\documentclass[twocolumn]{aastex631}
\usepackage{amsmath}
\usepackage{enumitem} 
\usepackage{soul}

\begin{document}
\title{Nature vs Nurture: Three Dimensional MHD Simulations of Misaligned Embedded Circum-Single Disks within an AGN Disk}
\shorttitle{Misaligned embedded mini-disks}
\author[0000-0003-0271-3429]{Bhupendra Mishra}\thanks{\href{mailto:mbhupe@camk.edu.pl}{mbhupe@camk.edu.pl}}
\affiliation{Nicolaus Copernicus Astronomical Center, Polish Academy of Sciences, Bartycka 18, 00–716 Warszawa, Poland}
\author[0000-0003-0271-3429]{Josh Calcino}\thanks{\href{mailto:jcalcino@mail.tsinghua.edu.cn}{jcalcino@mail.tsinghua.edu.cn}}
\affiliation{Department of Astronomy, Tsinghua University, 30 Shuangqing Rd, 100084 Beĳing, China}
\begin{abstract}
 Stellar mass black holes in the disks around active galactic nuclei (AGN) are promising sources for gravitational wave detections by LIGO/VIRGO. Recent studies suggest this environment fosters the formation and merger of binary black holes. Many of these studies often assumed a simple, laminar AGN disk without magnetic fields or turbulence. In this work, we present the first 3D magnetohydrodynamical simulations of circum-single disks around isolated and binary black holes in strongly magnetized, stratified accretion disks with turbulence driven by magneto-rotational instability.
 We simulated three scenarios with varying initial net-vertical magnetic field strengths: weak, intermediate, and strong. Our results show that weakly magnetized models produce circum-single disks aligned with the AGN disk's equatorial plane, similar to past hydrodynamic simulations. However, intermediate and strong magnetic fields result in randomly misaligned disks, contingent upon the availability of local ambient angular momentum within turbulent regions.
 Our findings emphasize the significant impact of ambient gas in the AGN disk on the inclination of circum-single disks, linked to magnetically induced inhomogeneity and angular momentum during disk formation. The presence of misaligned disks, both in single and binary black hole systems, could have profound implications for the long-term evolution of black hole spin and the inclination of the disk at the horizon scale.
\end{abstract}

\keywords{Active galactic nuclei (16); Magnetoydrodynamical simulations (1966); Black holes (162), Gravitational Waves (678)}
\section{Introduction}
\label{sec:intro}
Active Galactic Nuclei (AGN) are promising environments to facilitate the merger of binary black holes (BBHs), and may be the channel for observed GW events \citep{McK12,McK14,Stone17,McKernan24}. The progenitors that form the BBH system may arise due to in-situ formation \citep{Bellovary16,Secunda19,McK20b,dittmann2020,chen2023}, or be captured from the nuclear star cluster \citep{Morris93, Rasio04,Bartos17,Askar22,Leveque23}. Once inside the AGN disk, migrating BHs may be captured into migration traps \citep{Lyra10, Paardekooper10, Paardekooper11}, where close encounters can generated a bound binary through gas-assisted kinematic dissipation \citep{Tagawa.2020,jli2023, rowan2023,whitehead2023}, third-body interactions \citep{Leigh.2017,Samsing22,li2022}, or a combination thereof \citep{Secunda.2019, Secunda.2020}.

This channel is particularly attractive since it may explain some of the peculiarities seen in the observed GW population, such as progenitors within the pair-instability mass gap \citep[e.g. GW190521][]{gwtc21, Abbott20a}, and the near zero effective spin ($\chi_\textrm{eff}$) \citep{Callister21, abbot2021, McK2024}. In addition, given the gaseous environment of the AGN disk, there is the possibility that BBH mergers will lead to electromagnetic counterparts \citep{McK19a, Tagawa23}.

In recent years a significant focus has been placed on detailed study of companion-disk for embedded BBHs, predominately with 2D simulations \citep{Baruteau.2010, Li.2021, Li.2022, Rixin22, Rixin23, Rixin24}, with a handful of studies in 3D \citep{Dempsey22, whitehead2023, Dittmann24, calcino2024}. In general, these studies demonstrate that the torques due to the circum-single disk (CSD) around each progenitor BH play a dominant role in determining whether the binary will contract or expand \citep{Li.2021}. In 2D isothermal simulations, positive torques dominate leading to expansion. Adjusting the temperature profile and/or the equation of state tends to produce weaker CSD torques, resulting in contraction \citep{Rixin22}. The positive CSD torques are also diminished when the BBHs are embedded in a 3D AGN disk, although the shift from 2D to 3D leads to the contraction stalling, at least for BBHs on circular orbits \citep{Dempsey22, Dittmann24, calcino2024}.

One important shortcoming of these previous simulations is that they have been conducted assuming a smooth, non-turbulent background AGN disk. Turbulence in the AGN may arise at many different radial locations due to various mechanisms. Two of the most well known instabilities which could drive turbulence in astrophysical disks are the magneto-rotational instability (MRI) \citep{Balbus91, Pariev.2003, Lohnert22} and gravitational instability \citep{gammie2001, chen2023}. Little is known about the outer regions of AGN disks. The two well known AGN disk models \cite{Sirko.2003} and \cite{Thompson.2005} assume that marginally unstable region due to gravitational instability ($10^4-10^6$ Schwarzschild radius) could be stabilized due to heating source such as a stellar population. These outer regions are also relatively cold ($<10^3\,K$), which could mean that turbulence is not just MRI driven. However, in this study we assume that turbulence is MRI driven and its impact is primarily seen on the level of inhomogeneity developed in the flow. 
The outer regions of AGN disk however lack any global simulations due to timescale issue and therefore the effect of MRI is not well understood. 
The AGN disk modeling and different magnetic field configuration impact is beyond the scope of this study so we limit our model to weakly and strongly magnetized case only.

Detailed simulations of a disk with the interplay of MRI and gravitational instability \citep{Shlosman89} are still in developing stages \citep{Armitage01,Fromang04b,Fromang04c,Fromang04a,Zier23}. Using shearing box simulations with zero-net vertical magnetic flux \citet{Lohnert22} showed that gravitational instability itself can drive turbulence and later MRI could further amplify the available turbulent magnetic field (restricted to MRI suppression \citep{Pessah05} in strong field regimes). This interplay of MRI and gravitational instability behaves similar to $\alpha-\Omega$ dynamo and produces butterfly patterns (also seen in MRI only simulations) for toroidal magnetic field component. However, we limit our study by using gravitationally stable \citet{Thompson.2005} disk and its radial profile as our initial conditions for hydro variables. 
Local as well as global numerical simulations of strongly magnetized thin disks have been studied in detail \citep{Bai13, Salvesen16, Mishra20, Lancova19, Mishra22, Fragile23} and emphasized the need of strongly magnetized disks to prevent thermal, viscous and gravitational instabilities. All the previously simulated models in this context suggest a rapid magnetic field amplification if the setup is initialized with net-vertical magnetic field. In \citet{Salvesen16}, it has been shown that if one seeds a weak magnetic field of $\beta^\mathrm{mid}_0=10^4\,, 10^3$, the dynamo amplifies this magnetic field and gets a toroidal magnetic $\beta^\mathrm{mid}_0 \approx 0.1$ in accreting layers of the disk (where $\beta^\mathrm{mid}_0=P_\mathrm{gas}/P_\mathrm{mag}$ is initial midplane magnetic field parameter). In ideal MHD, an $\approx 10^5,\,10^4$ orders of magnitude of magnetic field amplification due to shear flow also makes the plasma compressible and develop large scale inhomogenities \citep{Salvesen16}. These large density inhomogenity play a key role in the evolution of circum-single disks in our study. 

In this paper, we consider the effects of magnetic fields and the generation of MRI driven turbulence in the evolution of embedded single and binary BHs. We produce MRI turbulent shearing-box simulations with varying levels of MRI turbulence by assuming different starting vertical magnetic field strengths, in a similar fashion to \citet{Salvesen16}. We then seed these turbulent shearing boxes with either single or binary BHs and study the birth and dynamics of the CSDs that form around each BH.

This paper is organized as follows: in Section \ref{sec:methods} we describe the numerical setup and different initial conditions to model shearing box simulations. In Section \ref{sec:results}, we describe the results for single and binary embedded black holes and the effects of turbulence of the AGN disk onto the disk inclination. In Section \ref{sec:discussion}, we discuss the astrophysical implications of such a misaligned circum-single disk formation and conclude our findings in Section \ref{sec:summary}.

\section{Setup}\label{sec:methods}
We generated our stratified local shearing box simulations with embedded BHs in four stages.
We list following four stages below in a sequence they are executed in order to model the embedded black holes in an AGN disk.
\begin{itemize}
    \item First simulating vertically stratified thin disk with uniform grid in a shearing box centered at a radial distance of $R_0 = 10^4\, G M/c^2$. This stage is similar to study performed by \citet{Salvesen16} and acts as a base model for developing a turbulent AGN disk before setting up the embedded black holes. We stop these first stage simulations once the magnetic energy reaches a saturation and no further field amplification due to MRI occurs. 
    \item In the second stage, we restart the simulations from the first stage and add nine levels of static mesh refinement (SMR). We evolve the turbulent disk simulation with SMR further to ensure no further fragmentation of gaseous structures developed compared to saturated MRI achieved in the first stage. The simulation also retains the same magnetic energy level as its earlier stage hence no further magnetic field amplification occurred and MRI maintained a saturated state achieved in the first stage.
    \item The third stage is to embed black hole with no general relativistic effects) at the center $(0,0,0)$ in the single black hole case and at a separation of (semi-major axis $a=0.25\,R_H$, where $R_H$ is Hill radius). Note that our binary orbit is circular and has zero inclination. In this specific stage the accretion into embedded black holes is not turned on (meaning no mass loss at the sink region). This is to ensure the numerical stability. Also, we grow the embedded black hole mass slowly such that it achieves its full mass at $t=0.5\,\Omega^{-1}_0$ (where $\Omega_0$ is the Keplarian orbital frequency at $R_0$)
    \item As the last stage of the simulation, we turn on the mass loss from the sink radius ($r_\mathrm{s} = 1.4\times 10^{-3}\,R_\mathrm{H}$) to allow the accretion onto each black hole and allow the circum-single disk evolution.
\end{itemize}
In the following subsections, we further describe the numerical details of each simulation stage. Note that these simulations are scale free with mass ratio of $(m/M) = 3\times 10^{-4}$.
\label{sec:setup}

\subsection{Vertically Stratified Turbulent Disk}
\label{sec:turbulent}
We use MHD version of Athena++ \citep{Stone.2020} to model the simulations reported in this article. The disk configuration uses \citet{Thompson.2005} radial profile with sheared and periodic boundary conditions in radial and vertical directions respectively. 
The key important physics we added in this study is effects of magnetic field and the MRI driven turbulence onto the evolution of embedded objects and disk formation around them. We evolve following set of MHD equations,

\begin{equation}
  \frac{\partial \rho}{\partial t} + \nabla\cdot (\rho \mathbf{v}) = 0 \,, 
  \label{eq:continuity}
\end{equation}
\begin{equation}
  \frac{\partial \rho\mathbf{v}}{\partial t} + \nabla\cdot\left(\rho\mathbf{v}^T\mathbf{v}
  + \mathbf{T} \right) =
  - \rho \left[2 \Omega \hat{\mathbf{z}} \times\mathbf{v}
  +2q\Omega^2 x \hat{\mathbf{x}} - \Omega^2 z \hat{\mathbf{z}}\right]\,, 
  \label{eq:eom} 
\end{equation}
\begin{equation}
  \frac{\partial \mathbf{B}}{\partial t} - \nabla \times \left(\mathbf{v}\times\mathbf{B}\right) = 0 \,, 
  \label{eq:induction}
\end{equation}
where $\hat{\mathbf{x}}$ refers to the radial direction,
$\rho$ is the mass density, $\mathbf{v}$ is the fluid velocity, $q = 3/2$ is the Keplerian
shear parameter, $\mathbf{B}$ is the magnetic field strength, and $\Omega = \Omega_0 \hat{\mathbf{z}}$ is the angular frequency at the center of the shearing box ($10^4\, G M/c^2$) which is unity in our simulations. 
The total stress tensor $\mathbf{T}$ is defined as
\begin{equation}
\mathbf{T} = \left(P+\frac{\mathbf{B}^T\mathbf{B}}{8\pi} \right)\mathbf{I} -
\frac{\mathbf{B}^T\mathbf{B}}{4\pi}
  \,
  \,,
  \label{eq:tensor}
\end{equation}
where $\mathbf{I}$ is the identity tensor, $P=\rho c_s^2$ is the gas pressure, and $c_s$ is the
isothermal sound speed ($c_s = 0.058$). The vertical density profile in the disk is initialized with a stratified disk in hydrostatic equilibrium,

\begin{equation}
    \rho_0(x,y,z) = \rho^\mathrm{mid}_0 \exp{\left(\frac{z^2}{2H^2}\right)}
\end{equation}
where $\rho^\mathrm{mid}_0$ is the initial midplane gas density and $H=c_s/\Omega_0$ is the initial disk scale height. We assume $\rho^\mathrm{mid}_0=1$, $\Omega_0=1$ and $c_s=0.058$ which gives us a disk scale height of $H=0.058$ (thin disk). The simulation domain spans $x=[-24H,24H]$, $y=[-24H,24H]$ and $z=[-4H,4H]$. The box is resolved with $(Nx,Ny,Nz)=(256,256,128)$ uniform grid in stage 1 of the simulation. The initial magnetic field is set as vertical \citep{Salvesen16} with magnetic field strength defined by $\beta^\mathrm{mid}_0=P_\mathrm{gas}/P_\mathrm{mag}$. The field configuration consists of vertical component of magnetic field as,
\begin{align}
    B_{x,0}(x,y,z) &= 0 , \\
    B_{y,0}(x,y,z) &= 0, \\
    B_{z,0}(x,y,z) &= B_0, 
\label{eq:bfield}
\end{align}
where, $B_0 = \sqrt{2 P_\mathrm{gas,0}/\beta^\mathrm{mid}_0}$. We used sheared periodic boundary condition in the x-direction and periodic boundary conditions in the y- and z-directions respectively. Note that simulations with net-vertical magnetic field and periodic vertical boundary conditions have caveats so we only simulated up to $\beta^\mathrm{mid}_0=100$ to avoid artefacts of winds launched and feedback in these simulations. We simulate three models with $\beta^\mathrm{mid}_0=10^4$ (weak), $\beta^\mathrm{mid}_0=10^3$ (intermediate) and $\beta^\mathrm{mid}_0=100$ (strong). The even smaller $\beta^\mathrm{mid}_0$ runs (e.g. reported in \citet{Salvesen16}, require mass injection and outflow boundary conditions in the vertical direction and therefore cannot be simulated with the current setup.) 
\begin{figure*}
\centering
\includegraphics[width=\textwidth]{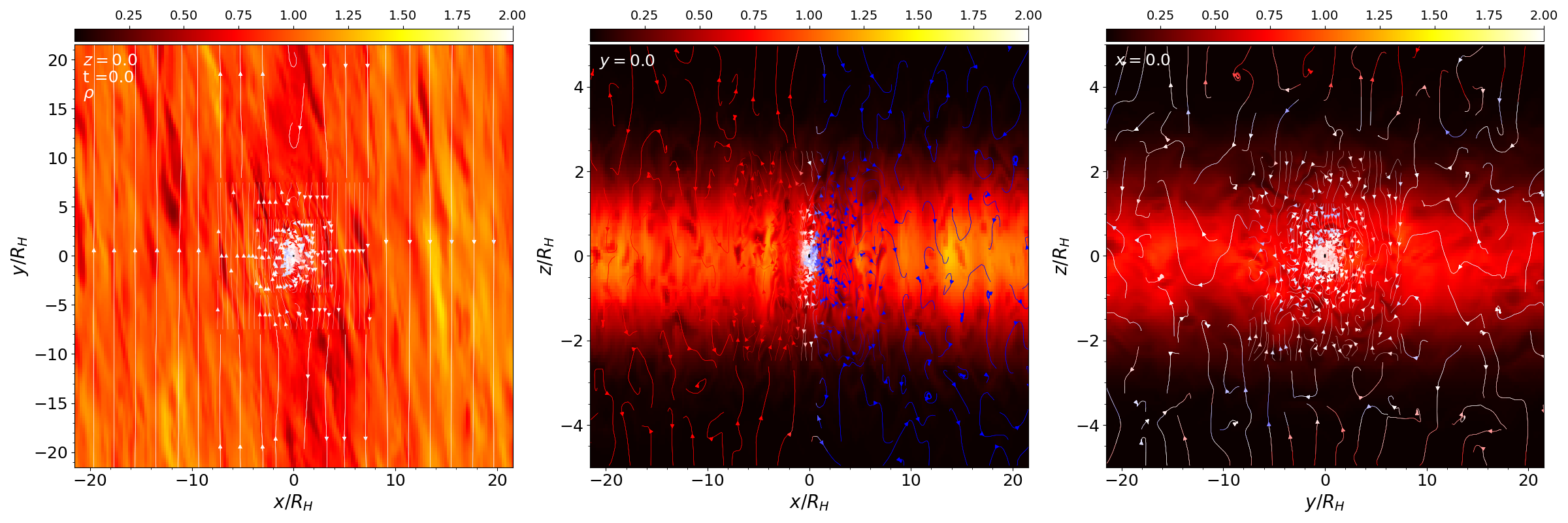}
\caption{Density slices at $t=200\,\Omega^{-1}_0$ for $\beta^\mathrm{mid}_0=10^4$ model. The axis labels are in units of Hill radius of the embedded black hole. The streamlines represent velocity field. The color of streamlines show the third out of plane velocity component. From left to the right, panels show mass density in the $x-y$, $x-z$ and $y-z$ planes respectively. These slices also coincide with the time when we seed the black hole. The high density of magnetic field streamlines is due to 9 levels of refinement on top of the base grid.}
\label{fig:turb}
\end{figure*}
\subsection{Turbulent Disk with Black Hole}
\label{sec:seedbh}
In the third stage of the simulation, we initialize a black hole (gravity term) at the center of the turbulent box (with a binary separation in case of two black holes). Prior to seeding the accreting object in our turbulent disk, we add SMR to resolve the sink region (which leads to nearly 400 grid points within sink radius). We ensure that MRI has achieved a saturated state before putting the black hole in the simulation. The black hole acts as a sink with spherically symmetric gravity. In order to have numerical stability, we evolve the black hole mass in such a way that it takes $t=0.5\,\Omega^{-1}_0$ for the black hole to grow its total mass chosen in this setup. The gravitational potential around the embedded black hole uses a spline function which is exactly Keplerian outside the softning radius $r_s$. Accretion flow that gets closer than softning radius feels softened potential. The mass loss from the sink region is treated using methods in \citet{Dempsey.2020}. As the accretion surface in these simulations is much larger than sink radius we ensure that angular momentum is not lost when we allow the mass loss. This is implemented as torque-free sink treatment in which during each time step, if a computational cell is within the sink radius of the black hole, and if the gas within that cell is bound to the black hole, we reduce the mass and change the velocity within the cell by 
\begin{equation}
    m^\prime = \frac{m}{1+\gamma n_b \Delta t}\,,
\end{equation}
and
\begin{equation}
    v^\prime = v+\left( \frac{\left( \gamma - \eta\right)n_b\Delta t}{1+\eta n_b \Delta t}\right)\left(\Delta v_\nu \hat{\mathbf{\nu}}+\Delta v_\Phi \hat{\mathbf{\Phi}}\right)\,,
\end{equation}
where $\gamma$ and $\eta$ dictate how much mass and angular momentum is lost to the black hole in this process. The $(\nu,\Phi)$ are the angles of a 3D spherical coordinate system centered on the black hole.
In our study we do not turn on the mass loss through the sink until $t=1.2\,\Omega^{-1}_0$. This is primarily motivated with keeping the angular momentum and mass loss in control when the disk is just starting to form from a turbulent accreting background gas. The magnetic field at the sink radius is left unchanged unlike treatment for the velocity and mass.

The second, third and fourth stage also consists of static mesh refinement (SMR) to resolve the sink region. In order to achieve this, we restart the turbulent disk model with uniform grid for zero cycle and turn on adaptive mesh refinement (AMR) feature of ATHENA++. Using the restart file from this turbulent run, we employ the refinement on existing run to further enhance the resolution in the central region and eventually resolve the circum-single disk formation. We restart the simulation with last restart for zero cycle but now we turn off the AMR and switch to SMR and let it evolve to make sure it achieves the same saturated MRI state as before adding any refinement. The final restart from saturated MRI run with SMR is used again to seed the black hole at its center. In the stages with black holes, we maintain the same level (9 SMR levels) of numerical resolution.

\begin{figure*}
\centering
\includegraphics[width=\textwidth]{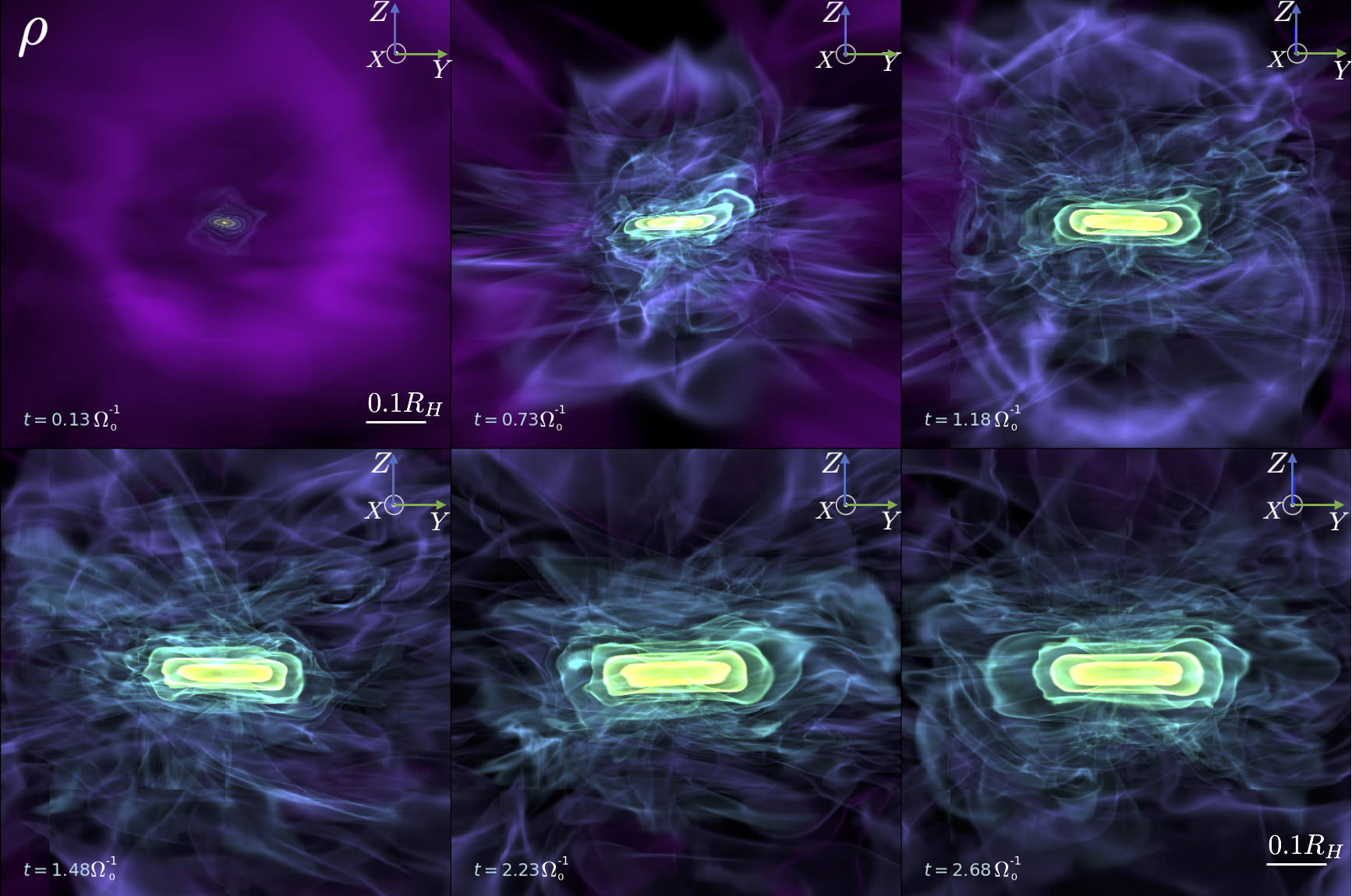}
\caption{Volume rendered density profiles for $\beta^\mathrm{mid}_0=10^4$ run at different times measured from its initial value when black hole is seeded. The colors show mass density with highest density at the central region. The disk forms in $x-y$ plane as shown by the axis triad in which z-axis is pointing out of the plane. The disk angular momentum is pointing towards z-axis (pointing out of the plane in these plots). The scale for these plots is shown at lower right corner of first and last panels. The image is zoomed in to show the circum-single disk region embedded in an AGN disk at much larger scale (see Section 2.1 for original simulation domain).} 
\label{fig:beta10k}
\end{figure*}

\section{Results}
\label{sec:results}
 In our simulations there are three different time scales: AGN orbital timescale ($\Omega^{-1}_0$), binary time scale (orbital period of the binary) and the smallest is dynamical time scale for each black hole. In the following section, we describe the results from each stage of the simulations and also further emphasize about the role of strong magnetic field in this study.
 
\subsection{Vertically Stratified Turbulent Disk}
\label{sec:results_turb}
In Fig. \ref{fig:turb}, we show three different planes of the turbulent disk by slicing at different coordinate slices. From left to right it shows density slices at $z=0.0$, $y=0.0$ and $x=0.0$ respectively. The streamlines show the fluid velocity field in the turbulent disk. The color of the streamlines show the out of the plane velocity component. The disk remains turbulent with eddies within the disk scale height (middle panel). The stratified disk starts with a midplane maximum density $\rho_0=1$ but as the turbulence develops, it forms slightly higher mean density structures. It has been already confirmed that stronger initial magnetic field strength leads to higher mean density clumps and more inhomogeneity \citep{Salvesen16} and we find the similar features in our simulations. Although, we start the simulation with a weak magnetic field of $\beta^\mathrm{mid}_0=10^4$, the shear flow develops a much stronger field by amplifying it via MRI. As we reduce the $\beta^\mathrm{mid}_0$ parameter, the turbulent structures seen in Fig. \ref{fig:turb} get more clumpy and achieve higher mean densities. However, we can not reduce the $\beta^\mathrm{mid}_0$ beyond $100$ in our study due to the magnetically driven winds in vertical direction and such a model will require mass injection and outflow boundary conditions in the z-direction. 

\subsection{Single Embedded Black Hole}
\label{sec:resultssinglebh}
The embedded single black hole in a turbulent AGN disk acts as a puncture and when we turn on the mass loss at the sink it begins to accrete the turbulent gas around it. The initial stages of the circum-single disk formation around such an embedded black hole begins with its orientation brought in by the turbulent blob accreted into the black hole. This initial in-fall from ambient turbulent AGN disk could set the disk orientation if deposited large enough inertia into the sink region. However, the inhomogeneity of ambient turbulent AGN disk is sensitive to initial magnetic field strength ($\beta^\mathrm{mid}_0$ parameter) and configuration. The weaker magnetic field case has smaller blobs of gas due to weaker eventual magnetic field after the amplification via MRI. However, when we increase the magnetic field strength (reduce the $\beta^\mathrm{mid}_0$ parameter), the size of the turbulent blobs begins to get larger and also achieves much higher densities. 

\begin{figure*}
\centering
\includegraphics[width=\textwidth]{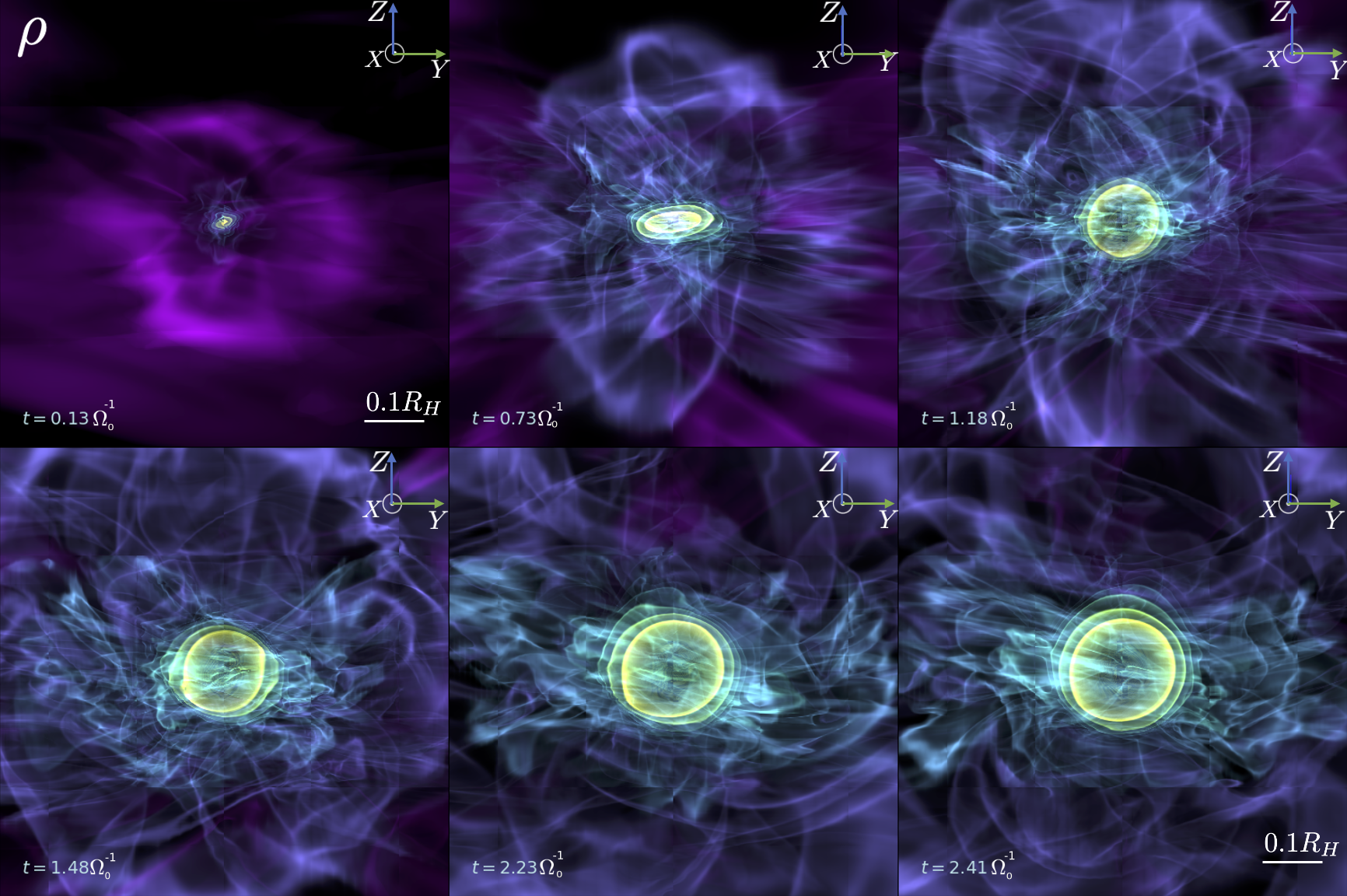}
\caption{Volume rendered density profiles for $\beta^\mathrm{mid}_0=10^3$ run at different times measured from its initial value when black hole is seeded. In the early times the disk undergoes a transient orientation with L pointing closer to z-direction ($t=0.73\,\Omega^{-1}_0$). At late times, the circum-single disk is aligned with y-z plane and pointing its total angular momentum vector in negative x-direction (into the paper). }
\label{fig:beta1k}
\end{figure*}

We observe the impact of inhomogeneity in a set of snapshots shown in Fig.~\ref{fig:beta10k} and Fig.~\ref{fig:beta1k}. Fig.~\ref{fig:beta10k} shows volume rendered density for the $\beta^\mathrm{mid}_0=10^4$ case which has the weakest seed magnetic field and also a weak eventual magnetic field amplification. The different panels in Fig.~\ref{fig:beta10k} show time measured in orbital periods ($2\pi/\Omega_0$). From left to right, both rows show a total of six snapshots and in all the panels the circum-single disk formation around the seeded black hole remains along the equatorial plane, which is also the AGN disk midplane. The final time $2.68\,\Omega^{-1}_0$ maintains the same disk orientation. We also found that once the disk has evolved for $\approx 1.0\,\Omega^{-1}_0$, its final orientation is already determined and further accretion from the turbulent surroundings does not alter its alignment. This is mainly due to the effects of already formed circum-single disk's high density and angular momentum. 

In Fig.~\ref{fig:beta1k}, we show the outcome of the intermediate magnetic field strength case. Similar to Fig.~\ref{fig:beta10k}, the snapshots show volume rendering of density for $\beta^\mathrm{mid}_0=10^3$ case. The stronger magnetic field case leads to a more inhomogeneous and clumpy AGN disk structure. The higher mean density gaseous blob when accreting into the embedded black hole, leads to misaligned circum-single disks with respect to AGN disk midplane. In Fig.~\ref{fig:beta1k} middle panel on top row shows the initial stage of the circum-single disk formation where the disk is aligned on the AGN disk plane but later when large inertia blobs bring in a different angular momentum from further away in the turbulent AGN disk, the mini-disk reorients with its net angular momentum pointing in the x-direction. As we keep evolving this setup longer, the disk orientation does not change and maintains its alignment in the y-z plane for this specific case. However, this disk orientation is mainly dictated by what time and location the black hole is seeded and what material is available to accrete at that specific space and time within the simulation domain. The initial accumulation of matter with certain angular momentum in the sink region starts to set the disk orientation and later grow the disk from inside out. 

\begin{figure*}
\centering
\includegraphics[width=\textwidth]{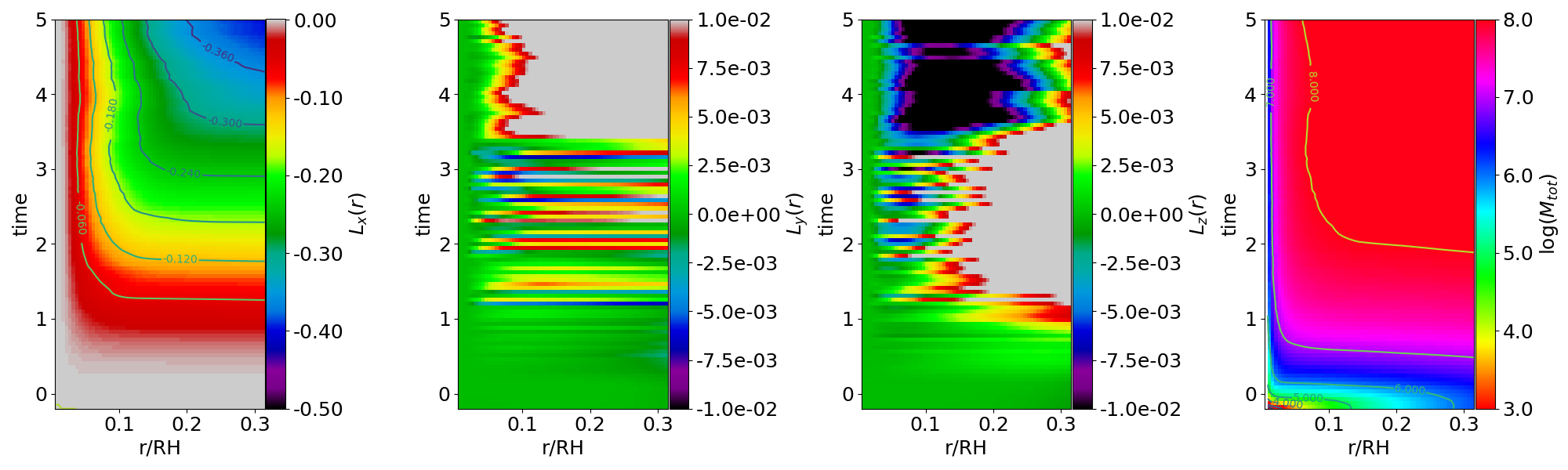}
\caption{The spherically averaged space-time ($\Omega^{-1}_0$) plot of the Cartesian components of the total angular momentum $L_x,L_y$ and $L_z$ respectively from left to right. This plot shows the growth of different $L$ components and eventual fate of the disk orientation. The rightmost plot shows the total mass growth around seeded black hole by measuring the spherically averaged integrated total mass as a function of spherical radius. All these panels are for our intermediate magnetic field strength run $\beta^\mathrm{mid}_0=10^3$.}  
\label{fig:angmom}
\end{figure*}

We further analyse the $\beta^\mathrm{mid}_0=10^3$ case to quantify its total angular momentum $L$ budget by computing the spherical average of each of its components. In Fig.~\ref{fig:angmom}, we show space-time plot of the magnitude of $L_i$ ($i=x,y,z$) as a function of time. We find that disk is primarily oriented on y-z plane with its total angular momentum vector pointing in negative x-direction. The $L_y$ and $L_z$ components remain negligible compared to the $L_x$ component for entire time duration. The rightmost panel shows the total mass build up around the embedded black hole. We notice a rapid circum-single disk formation and by $t\approx 1.0 \Omega^{-1}_0$, the disk as already formed up to a radius of $r\approx 0.2\,R_H$. The $L_x$ component maintains the same sign which represents a dense accretion disk and already determined disk orientation. On the other hand the remaining $L_y$ and $L_z$ components remain negligibly small yet having oscillatory behaviour in their sign. The sign fluctuations in $L_y$ and $L_z$ components is due to accretion of gaseous blobs from turbulent AGN disk. The small off-circum-single disk oriented angular momentum components are not affecting the existing circum-single disk orientation but rather begin to follow the disk rotation along its flow. 

\begin{figure*}
\centering
\includegraphics[width=0.32\textwidth]{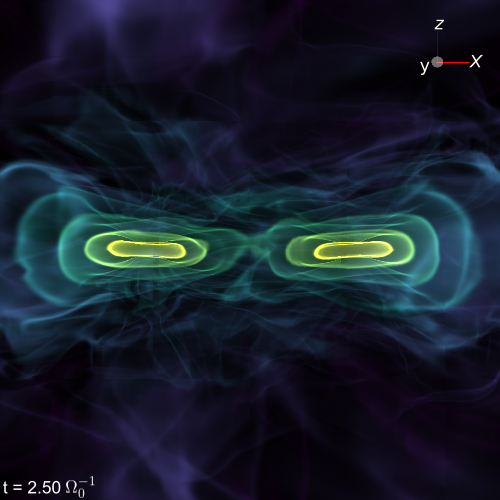}
\includegraphics[width=0.32\textwidth]{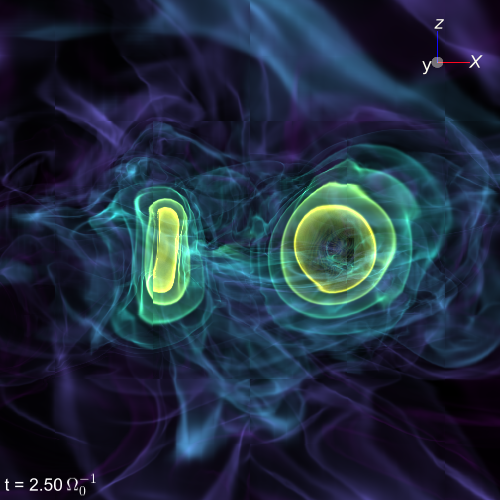}
\includegraphics[width=0.32\textwidth]{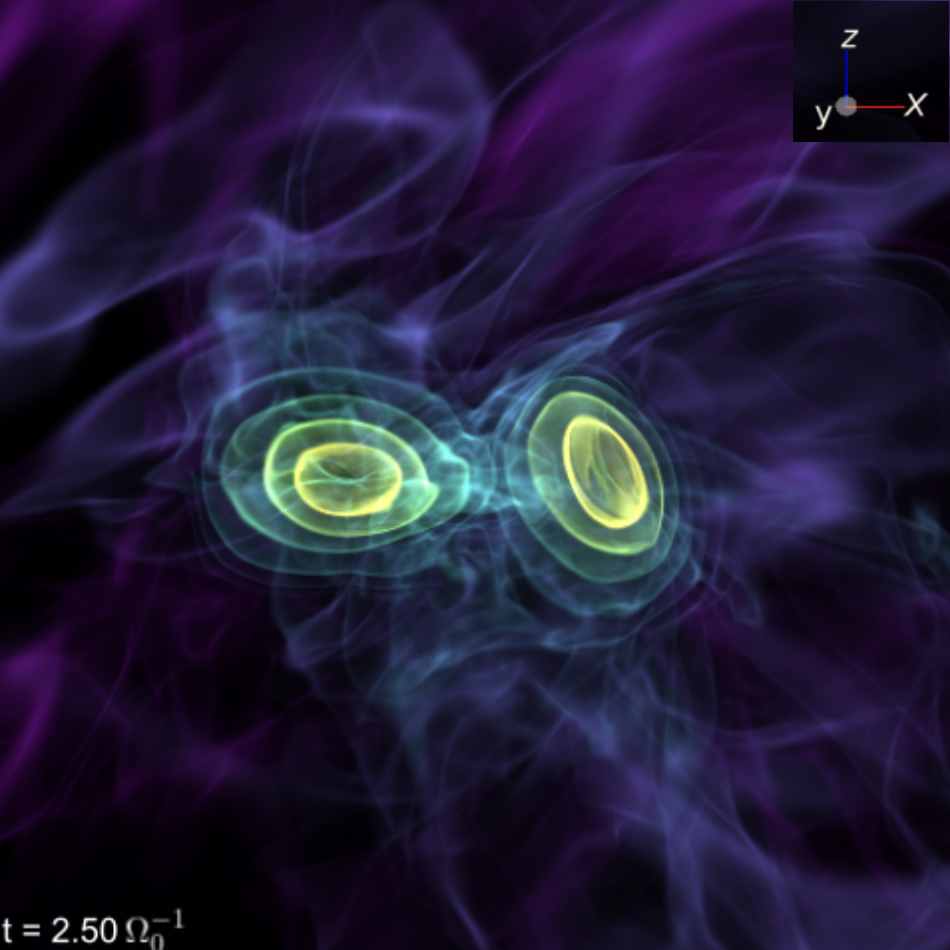}
\caption{The volume rendered density plots of an embedded binary system in a turbulent AGN disk. The three panels show weak, intermediate and strong magnetic field cases from left to right. The $\beta^\mathrm{mid}_0=10^4$ (left panel) has both circum-single disks aligned with the AGN disk angular momentum (z-axis). The intermediate magnetic field strength case ($\beta^\mathrm{mid}_0=10^3$) has one circum-single disk angular momentum aligned along y-axis whereas the other one along x-axis. In the strong magnetic field case ($\beta^\mathrm{mid}_0=100$), the circum-single disks are again carrying total net angular momentum, $L$ along two different directions. }
\label{fig:binary}
\end{figure*}

\begin{figure*}
\centering
\includegraphics[width=0.6\textwidth]{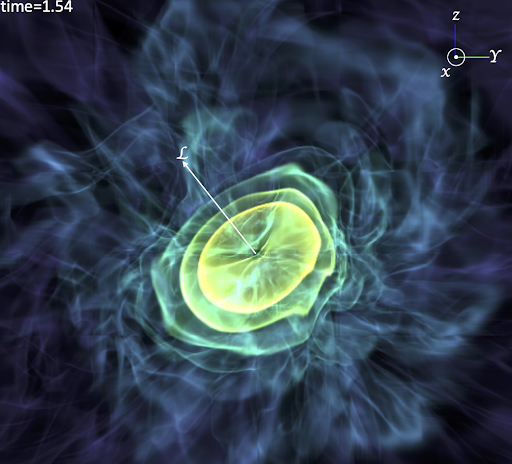}
\caption{The density plot (at $t=1.54\,\Omega^{-1}_0$) for the intermediate magnetic field strength case ($\beta^\mathrm{mid}_0=10^3$) but using eleven (9+2) levels of static mesh refinement and same base grid resolution as moderate resolution run to simulate the turbulence of the AGN disk. The SMR is extended to much larger Hill radius ($\leq 10\, R_\mathrm{H}$) and also enhance the effective resolution in the sink region. In this case the disk is not aligned to any specific coordinate axis unlike in the moderate resolution runs with the 9 levels of SMR. The disk angular momentum is about $40$ degrees from the z-axis. The seed time of the black hole, its mass growth and the time of beginning of the accretion remains same as moderate resolution run.}
\label{fig:hrimage}
\end{figure*}

\subsection{Embedded Binary Black Hole System}
\label{sec:resultsbinarybh}
Similar to the misaligned disk formation for single black hole, in stage 3 of our method, we initialize two black holes in a binary orbit and let them form circum-single disks around each. In Fig.~\ref{fig:binary}, we show three panels with weak to strong magnetic field from left to right at $t=2.50\,\Omega^{-1}_0$. The leftmost panel with $\beta^\mathrm{mid}_0=10^4$, retains the disk formation in the same plane as the AGN disk midplane. The intermediate and strong magnetic field cases ($\beta^\mathrm{mid}_0=10^3$ and $100$), form randomly oriented individual disks depending on the density and inertia of the turbulent gas at location and time the black holes are seeded. 
A stronger initial magnetic field leads to relatively larger inhomogeneities in the turbulent fluid. This suggests that the disk misalignement could undergo more transient fluctuations in the early stages of its formation. This is primarily due to the higher density and larger angular momentum of the accreted blobs created by the stronger magnetic field. In the intermediate field strength case the disk still forms with misalignment, albeit more slowly and with less transient fluctuations.

The circum-single disk around each embedded black hole attains its individual orientation due to initial phases of the accretion from turbulent ambient gas. The size of turbulent gaseous blobs and the time it is accreted could also impact the circum-single disk radius. If one of the CSD accreted a smaller gaseous blob whereas the other CSD accreted a gaseous blob with more mass and inertia, this could produce different initial size disks in the early phases of evolution. Eventually both embedded black holes form their own circum-single disks and remain in an initial prescribed circular orbit. 
In a right handed coordinate system, both circum-single disks are evolving similar to the disk evolution of a single black hole and form the disks from inside out.

\section{Discussion}
\label{sec:discussion}
\subsection{Grid Resolution vs Misalignment}
\label{sec:grideffects}
In the moderate resolution simulations reported here, we mainly used a total of 9 SMR levels on top of a base grid resolution. However, in all of our moderate resolution runs, we find the alignment of the circum-single disks along one of the coordinate axis (e.g. $\beta^\mathrm{mid}_0=10^3$ with disk angular momentum pointing in the negative x-direction). We further tested this by simulating different initial seed time of the embedded black hole after achieving saturated turbulence and found that even for the same initial magnetic field strength, the final orientation is along one of the coordinate axis and never along an off-coordinate axis. 
This is primarily due to a lack of extended SMR levels beyond the Hill radius and a moderate base grid resolution.
If we start the shearing box simulation with a uniform grid, the turbulence sets certain angular momentum for gaseous blobs. However, if the resolution is not high enough, these blobs tend to carry coordinate orientated total angular momentum. 

To test this, we ran even higher resolution simulations with 11 SMR levels (two higher than our standard simulations). The higher effective resolution does not qualitatively affect the saturated MRI amplified magnetic energy level or the turbulence level in the AGN disk. In order to distinguish if the effect is from the sink torque due to our boundary conditions or accretion from larger domain, the higher number of SMR levels are first used in the sink region only. However, the sink region with more refinement still showed the coordinate axis oriented circum-single disk formation. Therefore, we extended the SMR level to larger Hill radius ($\leq 10\, R_\mathrm{H}$) and found that the disk attains an orientation which is not aligned with any particular coordinate axis.

In Fig.~\ref{fig:hrimage}, we show one such simulation result by plotting the density iso-surfaces. The angular momentum vector of the disk in this high resolution run is not pointing along any specific coordinate axis direction but at about $40^\circ$ off the $z$-axis. 
This confirms the disk misalignment is not affected by the choice of a Cartesian coordinate system, but rather dictated by turbulence and the randomly accreted angular momentum.

The early evolution of disk around each black hole shows an off-axis misalignment (middle panel in first row of Fig.~\ref{fig:beta1k}). Such an off axis misalignment is due to high resolution central region being evolved before adding the sink particle. However, as the simulation evolves in time, the circum-single disk starts to accrete the plasma blobs away from the high resolution central region. This leads to the accretion of gaseous blobs with coordinate oriented angular momentum.

\subsection{Astrophysical Implications}
\label{sec:broadimpact}
Using these set of MHD simulations we find that the formation of circum-single disk around each progenitor black hole is very much dependent on the turbulence of the AGN disk. If the outskirts of an AGN disk (Toomre unstable region at about $10^5\, GM/c^2$) indeed has a population of stars and also presence of moderately strong magnetic field, the AGN disk could be significantly inhomogeneous \citep{Salvesen16}. The inhomogeneity will eventually lead to a formation of randomly orientated circum-single disk. 
If we take into account the amplification of weak available magnetic field to strong magnetic field and formation of misaligned disks around each black hole, this could play a key role in the long-term black hole spin evolution. Although the inner boundary of our circum-single disk is $1000\, r_g\, (Gm/c^2)$ for equal mass binary in circular orbit, the inner horizon scale disk might get torqued by this large scale misalignment of disk and lead to random black hole spin for each embedded progenitor. 

Recent BBH formation simulations have shown that retrograde binaries are preferentially formed \citep[e.g. see][]{jli2023, whitehead2023}, and merge much faster than their prograde counterparts \citep{Rixin22, calcino2024, Dittmann24}. However if hierarchical mergers are considered, \citet{McK2024} point out that retrograde mergers cannot make up the majority of mergers in the AGN disk. Post-merger BHs can have considerably large eccentricities around the central AGN, which can cause their accretion disks to become retrograde. For a prograde merger remnant, a retrograde accretion disk would reduce the remnant black hole spin. However a retrograde remnant would have their black hole spin increased, suggesting prograde mergers must be the dominant source. Randomly aligned CSDs around the progenitor and remnant BHs may act to reduce the prevalence of BH mergers with highly correlated spins. This in turn could help explain the LVK observations with low $\chi_\mathrm{eff}$ values \citep{Abbott20a, McK2024}. 

The misaligned disks are due to the turbulence level set by magnetized plasma. However, once the disk is forming around each progenitor, the magnetic field available from the turbulent AGN disk is further organized into strong toroidal magnetic field within each circum-single disk and achieves an amplification by $10^4$ times. Such a strong magnetic field could advect in to horizon scale disk and get further amplified leading to magnetically arrested disks around each of the progenitor and hence launches powerful jets from each embedded black hole \citep{Liska20, Mishra22}. The breakout radiation \citep{Tagawa23} from progenitor or post-merger black hole could then be observed \citep{Graham.2020}. This scenario could help in observing the electromagnetic counterparts from such an embedded binary black hole systems. 

The formation of CSDs around each progenitor also develops a toroidal magnetic field inheriting an opposite polarity compared to its binary companion. The typical plasma $\beta$ in these outer regions of the circum-single disks is of the order of few $100$ in our intermediate magnetic field case. Based on global isolated black hole accretion disk models by \citet{Zhu18, Mishra20, Lancova19, Mishra22}, we know that the disk midplane will have a current layer and opposite polarity of toroidal magnetic field above and below the circum-single disk midplane. Each circum-single disk with opposite toroidal field could set ideal scenario for magnetic reconnection. This could potentially lead to cascade of magnetic energy to smaller scale where the two circum-single disks interact periodically or aperiodically during tidal interaction. Such a region could cause particle acceleration to high energies and produce radiation as an observable source from these binary systems. Such a study has been reported in case of in-spiraling NS binary system \citep{Most22} but similar physics could play a role in the embedded binary black hole system in the AGN disk. 

\section{Summary}
\label{sec:summary}
We performed a set of 3D magneto-hydrodynamical simulations of single and binary black holes embedded in an MRI unstable AGN disk.
We tested three different magnetic field strengths with initial net-vertical magnetic field and $10^4, 10^3, 100$ $\beta^\mathrm{mid}_0$ values. The weakest magnetic field case behaves similar to 3D hydrodynamical simulations presented in other works \citep[e.g.][]{Dempsey22, calcino2024, Dittmann24}. However, the intermediate and strong magnetic field cases show randomly misaligned circum-single disks around each embedded black hole. We summarize the outcome of each model as follows: 
\begin{itemize}
    \item The single black hole cases show a circum-single disk formation along the equatorial plane of the AGN disk if the magnetic field is weak. However, the intermediate and strong magnetic field case shows the circum-single disk formation with random orientation. Note that similar to our hydrodynamical simulations, the circum-single disk around each progenitor forms torus like accretion structure as expected for a disk in equilibrium.

    \item Similar to the single black hole case, our binary black hole models with weak initial magnetic field shows both circum-single disk aligned with AGN disk midplane and the binary orbital plane. The intermediate and strong magnetic field case develops randomly misaligned disks around each black hole and leads to two different disk orientations for each of the black holes in the same binary system. 

    \item The disk misalignment is sensitive to when and where the black hole is seeded and due to what angular momentum is accreted onto each black hole during its early stages of the disk formation. Once the simulations have evolved for about $t=\Omega^{-1}_0$, the inclination of circum-single disk is stable and further accretion of randomly oriented angular momentum carrying blobs does not alter the disk alignment for the simulation duration we have evolved so far. 

    \item The disk orientation is sensitive to grid resolution. If the base grid resolution is not sufficient, the effects of Cartesian coordinate system far from central high resolution region cause the disks to align along certain coordinate axis direction. We confirmed that in case of high base grid resolution runs, this numerical artefact disappears. 

    \item The eventual fate of the progenitor black holes and their evolution is dependent on both nature (of accreting black holes and forming disks) and nurture (surrounding gaseous environment availed by AGN disk) to evolve the disks around each accretor. 
\end{itemize}
\section*{Acknowledgements}
This research is part of the project No. 2022/47/P/ST9/02315 co-funded by the National Science Centre and the European Union Framework Programme for Research and Innovation Horizon 2020 under the Marie Skłodowska-Curie grant agreement no. 945339. J.C. is supported by the National Natural Science Foundation of China under grant No. 12233004.
\bibliography{refs.bib}{}
\bibliographystyle{aasjournal}

\end{document}